# Designing Explainable AI for Healthcare Reviews: Guidance on Adoption and Trust

Eman Alamoudi
*School of Computing*
*Newcastle University*
*Newcastle, UK*
Department of Information Technology
College of Computers and Information Technology
Taif University, Taif 21944, Saudi Arabia
E.S.O.Alamoudi2@newcastle.ac.uk

Ellis Solaiman
*School of Computing*
*Newcastle University*
*Newcastle, UK*
Ellis.Solaiman@newcastle.ac.uk

*Abstract*— **Patients increasingly rely on online reviews when choosing healthcare providers, yet the sheer volume of these reviews can hinder effective decision-making. This paper summarises a mixed-methods study aimed at evaluating a proposed explainable AI system that analyses patient reviews and provides transparent explanations for its outputs. The survey (N=60) indicated broad optimism regarding usefulness (≈82% agreed it saves time; ≈78% that it highlights essentials), alongside strong demand for explainability (≈84% considered it important to understand why a review is classified; ≈82% said explanations would increase trust). Around 45% preferred combined text-and-visual explanations. Thematic analysis of open-ended survey responses revealed core requirements such as accuracy, clarity/simplicity, responsiveness, data credibility, and unbiased processing. In addition, interviews with AI experts provided deeper qualitative insights, highlighting technical considerations and potential challenges for different explanation methods. Drawing on TAM and trust in automation, the findings suggest that high perceived usefulness and transparent explanations promote adoption, whereas complexity and inaccuracy hinder it. This paper contributes actionable design guidance for layered, audience-aware explanations in healthcare review systems.**

## I. INTRODUCTION

Online patient reviews have become a significant resource for people making healthcare decisions. Patients often read others' experiences with doctors, clinics, or hospitals to gauge quality of care, bedside manner, wait times, and other factors. However, popular providers can have hundreds of reviews, making it time-consuming to distill overall sentiment or specific pros and cons [1]. Moreover, such analysis holds substantial value for healthcare stakeholders, including hospital administrators, policymakers, and the broader health sector, by offering data-driven insights into service quality, identifying recurring issues, and supporting evidence-based improvements in healthcare delivery [2]. Automated sentiment analysis and summarization of healthcare reviews is a promising approach to manage this information overload. By leveraging Natural Language Processing (NLP), such a system can extract overall positive/negative sentiment trends and highlight key aspects mentioned across reviews (e.g. cleanliness, staff attitude, treatment effectiveness). Crucially, explainability is needed in this context: users must trust the AI's analysis, which in turn requires understanding the reasoning behind the AI's conclusions [3]. Explainable AI (XAI) refers to methods that make an AI system's operations and outputs understandable to humans [4]. In the case of sentiment analysis, explainability might involve showing which words or phrases in a review led to a positive or negative classification, or providing a plain-language rationale for the summary.

The investigation is framed by insights drawn from research on technology adoption and trust. The Technology Acceptance Model (TAM) posits that two primary factors – perceived usefulness and perceived ease of use – determine users' attitudes towards a new technology, influencing their intention to use it [5]. If users believe an explainable review analysis system will genuinely help them (for example, by saving time or improving decision quality) and is easy to interact with, they are more likely to embrace it. Trust in automation is another key consideration: users must trust that the AI system is accurate, reliable, and acting in their best interest. Prior research indicates that an AI system's accuracy has a significant effect on people's trust, and lack of trust can lead to disuse of the technology [6]. Conversely, providing transparent explanations for AI outputs can enhance trust by making the AI's decision process more understandable [7]. These perspectives from TAM and trust in automation literature align with broader trends: people tend to prefer technologies they find useful, easy, and transparent, and they often reject "black-box" AI systems that do not provide clear justifications [8]. Indeed, explainability in AI has been shown to positively impact user trust and perceived usefulness [9]. At the same time, excessive complexity or technical jargon in explanations can undermine trust, especially for non-expert users [10]. With the increasing reliance of patients on online reviews when selecting healthcare providers, the need has emerged for tools capable of summarising these reviews and presenting their outcomes in a transparent manner. While explainable AI has been explored in domains such as e-commerce, where it is used to clarify recommendation logic or highlight patterns in product reviews, the healthcare context presents a more sensitive and consequential setting. Unlike consumer choice, decisions about healthcare providers are closely tied to personal well-being and the quality of care received. Therefore, patient-facing explanation tools must not only summarise content effectively but also tailor their explanatory methods to support clear understanding and foster trust. This study contributes to one of the first systematic evaluations of adoption and trust in explainable review summarisation within the healthcare context, addressing a gap that has not been adequately examined in prior work. This shifts the focus from purely technical optimisation toward understanding real-world adoption factors, particularly how explanation delivery shapes user confidence and decision-making.

**Contributions.** This paper complements our prior technical work by shifting from building explainable review-analysis pipelines to assessing whether, and under what conditions, users would adopt and trust such tools. We (i) quantify perceived usefulness, trust drivers, and explanation preferences (N=60); (ii) summarise expert perspectives on XAI methods/risks; and (iii) distil a design checklist for layered, audience-aware explanations and data-credibility cues, grounded in TAM and trust-in-automation.

The remainder of this paper is organised as follows. The related work section reviews existing research on explainability within healthcare contexts. The methodology and analysis section then describes the study design and analytical procedures. The research questions section outlines the questions guiding this study. The results section presents the empirical findings, followed by a discussion that interprets them in the context of existing research. The paper then outlines future work, acknowledges the study's limitations, offers recommendations, and concludes with the final remarks.

## II. RELATED WORK

Prior studies on online patient reviews demonstrate their influence on healthcare decision-making [11]. In [12] authors found that awareness and reliance on physician rating sites has increased steadily and reviews directly shape trust and choice. However, the difficulties processing large volumes of heterogeneous comments, reinforcing the need for summarisation technologies [13].

Explainable AI (XAI) methods provide interpretability across clinical and non-clinical tasks; prior work highlights that the effect of explanations on trust depends on model accuracy, task, and audience [14]. While healthcare XAI is expanding, existing evidence shows explanations can both increase and decrease trust depending on context [14]; underscoring the need to study adoption and trust for explainable review tools. This paper fills that gap by combining quantitative and qualitative user research to identify user expectations and adoption barriers.

## III. METHODOLOGY AND ANALYSIS

The study adopted a mixed-methods design [15] that integrated both quantitative and qualitative components. The first component consisted of an online survey designed to capture a broad range of data, including quantitative measures of participants' attitudes as well as qualitative reflections on their experiences. The survey was structured into four sections: (i) demographic information and participants' background in artificial intelligence (AI) and natural language processing (NLP); (ii) current practices in consulting online reviews; (iii) perceptions of the proposed explainable sentiment analysis system, assessed through Likert-scale items measuring usefulness, trust, and preferred explanation formats; and (iv) Open-ended questions targeting facilitators and barriers to adoption of a proposed explainable patient review system. The survey consisted of 18 questions, along with an item for comments and an item for volunteering in the follow-up interview, and was provided in Arabic and English to ensure accessibility and to meet participants' language preferences. A total of sixty respondents completed the survey.

The second component focused on collecting in-depth qualitative data through an online follow-up interview [16] consisting of four open-ended questions, directed at participants with technical expertise in artificial intelligence. These questions were specifically oriented toward the technical dimensions of explainability in AI systems, with particular attention to how explanations are generated, adapted, and interpreted in healthcare-related contexts. This interview component was treated as an exploratory stage aimed at surfacing key technical priorities that may be examined more systematically in subsequent, larger-scale studies, and the resulting insights were used to enrich the survey findings with informed expert perspectives on the interpretability of AI systems. Survey data were summarised with descriptive statistics; Likert items are reported as percentages and means. Associations were tested using Pearson's r (or Spearman's ρ if non-normal). Open-ended responses and interview notes were thematically analysed. This combined approach provided general trends alongside deeper qualitative insights, integrating user attitudes with expert perspectives. Microsoft Forms was used to administer both the survey and the interview. Ethical approval was obtained in line with institutional requirements, and identifying information was excluded from the analysis to preserve anonymity.

The survey questions and interview protocol used in this study can be accessed via the following link: https://doi.org/10.5281/zenodo.17229668

## IV. RESEARCH QUESTIONS (RQ)

The study was guided by six research questions (RQs): **RQ1** examined user behaviour and needs, particularly how healthcare reviews are read and what difficulties users face. **RQ2** focused on perceived usefulness: whether users see value in AI-assisted review analysis and in what form. **RQ3** explored the role of explainability in shaping trust, addressing whether transparent reasoning encourages acceptance. **RQ4** investigated preferred explanation formats. **RQ5** looked at adoption enablers and barriers. **RQ6** explored expert perspectives on XAI methods and challenges.

## V. RESULTS

### A. Survey

Quantitative responses were statistically analysed, and the key results are summarised as follows.

***Participant Profile*:** Among the sixty participants were mostly adults aged 25–44, representing 83% of the sample (50% in the 35–44 group and 33% in the 25–34 group), while 8% were

45–54, 5% were 18–24, and a small proportion (3%) were 55 or older. As shown in Fig. 1, which illustrates the distribution of respondents' education levels, education levels were notably high: 95% held a university degree (47% Master's, 28% Bachelor's, and 20% PhD). Additionally, 70% reported intermediate or advanced knowledge of AI, while 37% had intermediate and 15% advanced expertise in natural language processing (NLP) specifically. In terms of prior exposure, 67% had previously interacted with AI-based systems in daily life, 17% were unsure (likely due to AI features operating in the background of apps), and another 17% had not knowingly engaged with AI. Together, these findings reflect a mix of technical backgrounds and a clear representation of digitally literate early adopters.

*Use of Reviews (RQ1):* Consulting reviews was widespread: 60% said they always read patient reviews before selecting a healthcare provider, 28% sometimes, and only 12% rarely or never. This distribution, illustrated in Fig. 2 and depicting how frequently participants consult healthcare reviews, shows a strong tendency towards integrating patient feedback into decision-making, with comparatively few participants disregarding such reviews. The reviews were not only consulted but trusted; on a five-point scale, among the 53 participants who read reviews, the mean helpfulness rating was 4.32, with no respondent rating them below neutral. As illustrated in Fig. 3, which presents the different reading approaches participants use when engaging with healthcare reviews, respondents' strategies varied: 43% skim to extract the gist, 36% read reviews in full, and 21% search for keywords. In terms of time, 53% typically spend 4–10 minutes reading, 23% spend only 1–3 minutes, and 28% spend more than 10 minutes, as depicted in Fig. 4, which shows the distribution of reading durations. These findings reveal a common pain point: reviews are valued but time-consuming, and many users resort to shortcuts.

*Perceived Usefulness (RQ2):* Perceptions of the proposed AI system were highly positive. As demonstrated in Fig. 5, which presents participants' ratings of the system's perceived usefulness, 82% agreed or strongly agreed that it would save time, while only 3% disagreed. 78% agreed it would surface essentials without requiring full reading, and 75% endorsed aspect-based organisation, which would group comments under themes such as waiting time or cleanliness. Participants also recognised more abstract benefits: 70% believed the system could make decisions more reliable, and 68% thought it could reduce bias through standardised analysis. The lack of strong disagreement across all statements is noteworthy; even the most sceptical respondents were neutral rather than opposed.

*Explainability and Trust (RQ3):* Participants placed a premium on transparency. On a five-point scale, the importance of understanding why a review was classified as positive or negative had a mean of 4.3, with 70% selecting the maximum score. Similarly, when asked if explanations would increase trust, the mean was 4.4, with 82% selecting the top two categories. There was a strong positive association between the two items (Pearson's $r = 0.808$, $p < 0.001$); participants who rated explanations as essential also reported higher expected trust.

*Preferred Explanation Formats (RQ4):* Preferences varied but a plurality (45%) chose a mixed mode, combining text with visuals or keywords. Twenty-seven per cent preferred text only, favouring concise natural-language rationales. Twenty per cent preferred highlighted keywords, and only 8% favoured graphs alone, as shown in Fig. 6, which illustrates the distribution of explanation format preferences. In total, 72% demanded textual explanations in some form. This reinforces the need for layered, multi-modal explanations that combine the clarity of words with the immediacy of visuals.

*Prior Use of Tools:* A striking 85% had never used a similar system, confirming the novelty of explainable AI in healthcare reviews. Those with experience cited tools for detecting fake reviews in e-commerce, but none mentioned healthcare.

To complement the quantitative findings, the open-ended responses were thematically analysed to address **RQ5**. Participants described adoption enablers in Question 17 (Q17) and adoption barriers in Question 18 (Q18), with additional comments provided in the final question. The analysis revealed several key themes.

*Usability and Simplicity:* Usability emerged as the most frequently cited expectation among participants, with 18 respondents (30%) in Q17 emphasising the importance of an intuitive, accessible, and straightforward design. Phrases such as "easy to use", "user-friendly", "easy to navigate", "easy user interface", and "clarity and simplicity" (translated from Arabic: "الوضوح والبساطة") reflected a strong preference for systems requiring minimal learning effort while remaining inclusive for non-expert users. Notably, one participant stated that the system should be understandable even for "non-technical users, like my mother", underscoring that usability is not only a matter of convenience but also of equity and accessibility.

In contrast, the absence of simplicity was consistently framed as a serious deterrent. Approximately 17 respondents (28%) in Q18 cautioned against "complexity" (translated from Arabic: "التعقيد") or "difficulty of use" (translated from Arabic: "صعوبة الاستخدام"). Others noted that "too much text and steps" or a "complex interface" would render the system more burdensome than beneficial. As one participant explained, if the system is "complex and harder than skimming reviews myself", it effectively nullifies its intended purpose.

*Accuracy and Reliability:* Although 10 participants (17%) in Q17 explicitly emphasised the need for "Accurate results" or "accurate answers" (translated from Arabic: "دقيق في الاجابات"), accuracy became even more decisive in Q18. Here, 18 participants (30%) rejected systems they perceived as inaccurate, using phrases such as "Unreliable results", "عدم دقة المعلومات" (lack of accuracy in information), "إذا لم يكن دقيقًا وواضحًا" (if it is not accurate and clear), and "غير دقيق" (not accurate). Others linked this concern to system performance, noting "عدم الدقة في النتائج" (inaccuracy in results).

*Explainability and Transparency:* Explainability emerged as a notable theme, with 8 participants (13%) in Q17 explicitly requesting features such as "transparency and ability of explanation" (translated from Arabic: "الشفافية وقابلية التفسير"). Participants consistently emphasised the importance of being provided with "clear explanations". For example, one respondent described "Reasonable explanations" as a feature that would make the system appealing, where "reasonable" was interpreted as explanations that credibly align with the review content rather than offering generic or far-fetched justifications. Another participant wrote that the system should "simple explanations", signalling that the AI should not only produce results but also accompany them with an easy-to-grasp rationale. More specific suggestions highlighted the value of "clear visual explanations of sentiment (e.g., color-coded summaries or sentiment graphs)", "transparency about how the analysis is performed", and "clear explanation of why a review is classified as positive, negative, or neutral (highlighting key phrases or patterns)." Others stressed the importance of "transparency about how the AI model works" and the ability to trace back to the original review text, thereby ensuring traceability. By contrast, 4 participants (7%) in Q18 framed opacity as a deterrent. Concerns included "The explanation being too technical", "Lack of transparency about how conclusions are reached", and "Overly technical explanations that are hard for non-experts to understand." One participant also criticised "over-reliance on automation without an option to see the original text", indicating that lack of transparency not only obscures understanding but also removes users' ability to verify results independently.

*Speed and Efficiency:* Responsiveness emerged as an important expectation, with 8 participants (13%) in Q17 explicitly requesting features such as "Fast," "quick response," and the Arabic phrase "سريع الاستجابة" (tfast response). For these users, efficiency was closely tied to the system's core value proposition: reducing the time and cognitive effort otherwise required to manually process numerous reviews. Conversely, when efficiency was absent, speed transformed from an enabler to a deterrent. In Q18, 8 participants (13%) highlighted "البطء" or "بطيء الاستجابة" (slowness or slow to respond) as discouraging factors. Their responses underscored that delays undermine the very promise of automation. As one participant observed, a slow system would "defeat the benefit of automation."

*Data Credibility and Source Quality:* Concerns regarding the quality of input data were particularly salient across participants' responses. In Q17, 5 participants (8%) explicitly emphasised the importance of credibility. For example, one respondent wrote "الجوده فى كل شي" (quality in everything), while another stated that the system should "provide credibility." These responses reveal an acute awareness of the well-known problem of "garbage in, garbage out", whereby the reliability of automated outputs is inherently constrained by the reliability of the underlying data. Several participants further proposed specific measures such as ensuring "transparency about where its data comes from," enabling the system to "differentiate true reviews from spam or paid reviewers," and the necessity to "استبعاد التعليقات الوهمية" (exclude fake comments). Such comments underscore the expectation that the system should not indiscriminately treat all reviews as equally valid, but instead provide visibility into data provenance and filtering mechanisms. In Q18, these concerns became even more pronounced. About 5 participants (8%) expressed clear concerns regarding the credibility of reviews. One participant highlighted the issue of "مراجعات غير صحيحة كتبت للتسويق أو الضرر" (reviews written for marketing purposes or to cause harm), while another described such cases as "fake/marketing comments"). These responses indicate that data authenticity plays a direct role in shaping user trust in the system. Additionally, one participant summarised this concern succinctly: "If reviews are fake, they will become a barrier to adopting the system."

*Privacy and Data Security:* Privacy, although less frequently cited, emerged as a noteworthy consideration in both Q17 and Q18, albeit with different emphases. In Q17, 2 participants explicitly framed privacy as a desirable system feature. One suggested "No need for logins", highlighting a preference for unrestricted access without creating accounts, while another emphasised "ensure privacy" (translated from Arabic: "ضمان الخصوصية"). These responses position privacy as a positive enabler of usability, lowering barriers to entry and reassuring users that their engagement with the system would not expose them to unnecessary risks. By contrast, in Q18 privacy appeared as a source of concern. Around 3 respondents (5%) expressed apprehension about vulnerabilities such as "weak privacy protection", "Data security and how my inputs are used", and "Having personal information." Here, privacy was framed not as an optional convenience but as a potential liability that could actively discourage use.

*Secondary Themes*: Beyond usability, accuracy, and explainability, participants also identified secondary themes that, though less frequent, highlight additional factors influencing adoption.

- *Visualisation*: Visualisation was raised positively in Q17 by 6 participants (10%), who requested features such as "graphical charts", "color-coded summaries", or "provide clear visual explanations of sentiment (e.g., color-coded summaries or sentiment graphs)." These responses indicate that visuals were perceived as enhancing clarity and engagement. However, in Q18 no participant explicitly cited the absence of visualisation as a barrier.

- *Bias and Objectivity*: Concerns about bias were noted across both questions by approximately 5 participants (8%). In Q17, users called for objectivity, warning against "تحيز" (bias). In Q18, bias was described more explicitly as a deterrent: some respondents cautioned against the system "Being biased or has some advertisements" or suffering from the "Inherent bias of the data because of its source." Advertisements and sponsorship were viewed as particularly problematic, signalling commercial influence.

- *Information Overload*: information overload was flagged as a discouragement in Q18 by 5 respondents (8%). Phrases included "تحليلات طويلة" (long analyses), "too much text", and "Provides information that I'm not willing to see."

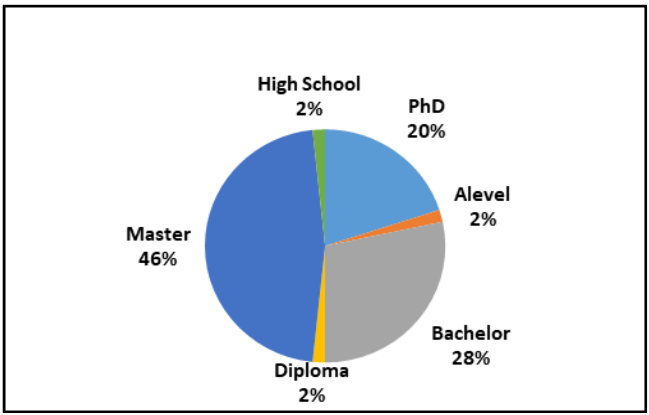

Fig. 1. The respondent's education level.

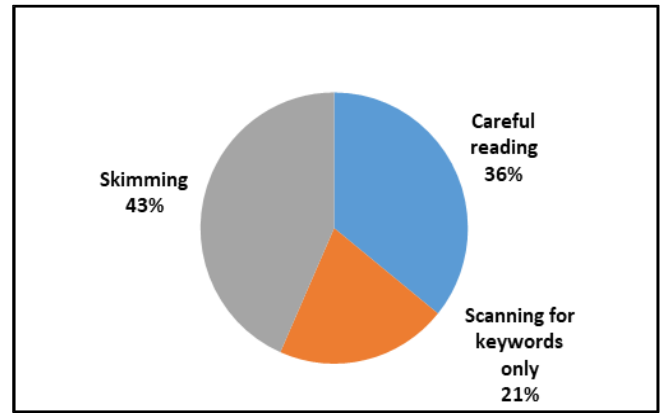

Fig. 3. Participants' reading styles of healthcare reviews.

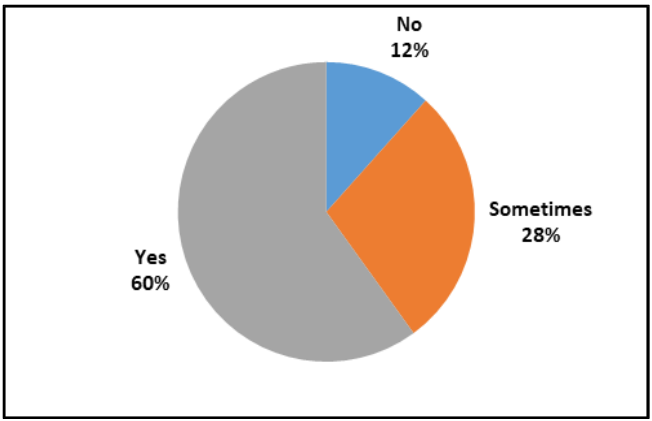

Fig. 2. Proportion of participants who read healthcare reviews.

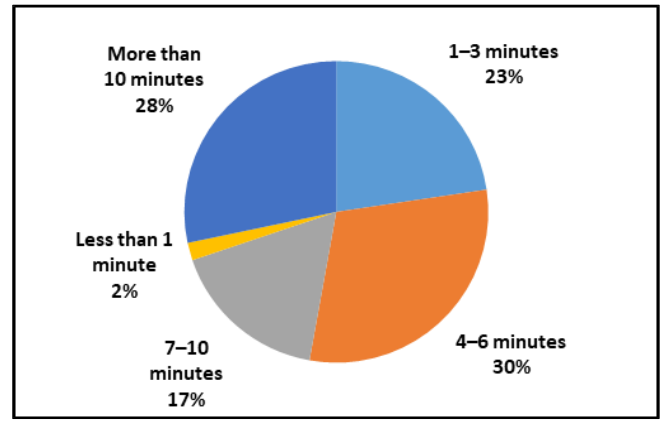

Fig. 4. Participants' distribution by time spent reading healthcare reviews.

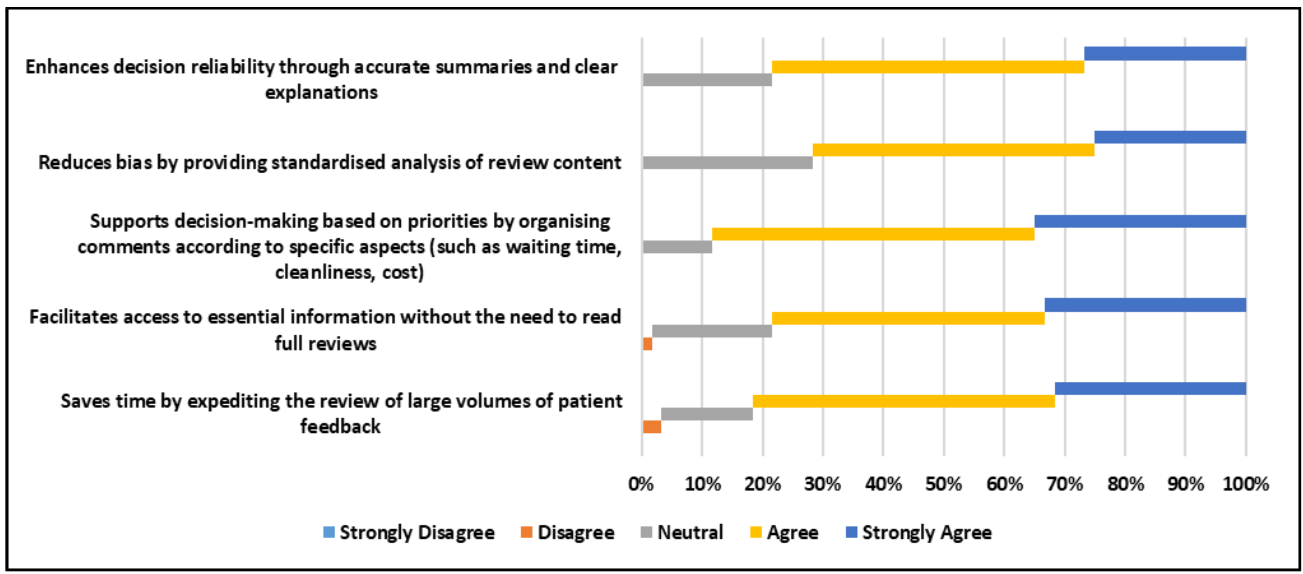

Fig. 5. Reported benefits of an XAI-assisted summarisation system.

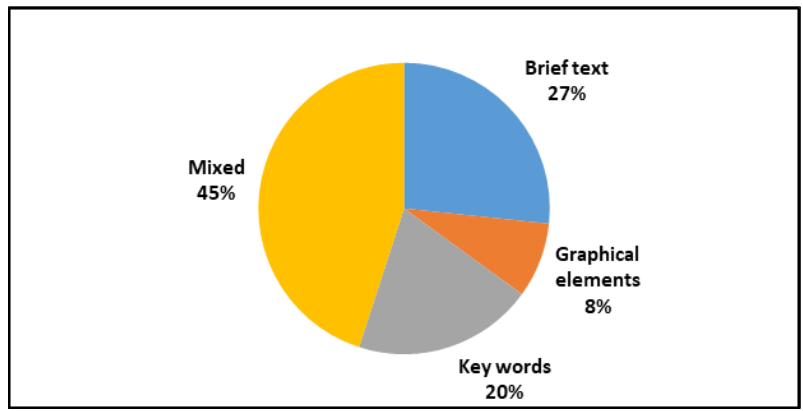

Fig. 6. Preferred explanation formats.

*Additional Comments:* In the final question, which invited participants to provide any further remarks or suggestions, 13 answers were given, most of which were positive and encouraging (e.g., "Great idea," "Wish you success"), with some noting the potential benefits of the system in the healthcare sector.

*B. Interview*

To address **RQ6**, we conducted interviews with five participants, using four predefined questions focused on explainable AI (XAI) methods and challenges. The analysis below summarises their responses, capturing both technical and user-facing considerations.

Participant 1 noted that a major obstacle with current XAI techniques arises when explanations are vague (unhelpful) or not easy to understand. They stressed that overly long or complex outputs hinder comprehension. To make explanations user-appropriate, they proposed a layered approach: beginning with a simple summary and enabling users to explore more detail as needed. They added that visual elements, such as charts or icons, can help support understanding and reduce cognitive effort.

Participant 2 noted that XAI explanations may not be clear or accurate and should be tailored to different audiences: detailed numbers for experts and simple language, visuals, or examples for non-specialists. They also highlighted feature ablation as a practical way to illustrate how changing or removing a single input influences the model's output. However, regarding future challenges, they cautioned that multi-stage XAI pipelines "could be slower or need more computing resources, especially with large datasets or complex models".

Participant 3 underlined the difficulty of balancing model performance with interpretability. They noted that, while transparent models such as decision trees remain a favourite for their simplicity, they often underperform on complex datasets. In contrast, post-hoc methods like SHAP and LIME can achieve higher performance but introduce considerable computational overhead, which can limit scalability in practice. They emphasised that explanations should reflect the model's actual behaviour rather than merely appear plausible to users, stating that explanations should "not just seem to be right to humans," underscoring the importance of fidelity. They also highlighted the importance of adapting explanations to different audiences, with more technical detail suited to expert users and simplified or visual formats likely to be more appropriate for non-specialists. To address performance issues, particularly with unstructured data, they proposed strategies such as clustering processing tasks, optimising storage configurations, and handling computation in stages. They further noted that incorporating end-user feedback can help improve responsiveness, usability, and overall experience.

Participant 4 highlighted the trade-off between model accuracy and interpretability, noting that methods such as SHAP and LIME, though powerful, can be costly to compute and sometimes unstable with large datasets. They stressed the need to tailor explanations by providing technical detail for experts (e.g., feature attributions, model internals, or visual/interactive representations) while offering short, plain-language summaries for non-specialists. They described SHAP as one of the more interpretable techniques, but noted that combining multiple stages within the explanation pipeline may introduce additional computational overhead, which can limit scalability in real-world healthcare settings.

Participant 5 provided a comparative and critical perspective on current XAI practices. They highlighted that "a single explanation method rarely satisfies all needs", emphasising the persistent tension between faithfulness and interpretability. According to their view, simple surrogate models can sometimes oversimplify or even distort the underlying model behaviour, whereas more faithful approaches (e.g., influence-based techniques) may be too technically demanding for non-expert users. They also noted that local explanations often do not accumulate into a coherent global understanding of model logic and expressed concerns about computational and latency overhead when explanation methods are deployed in real-time settings. They further suggested that explanation strategies should be adapted to user profiles by developing persona-oriented explanation templates. For non-technical users specifically, they recommended prioritising example-based and counterfactual explanations that demonstrate how changes in input might alter an outcome, supported by everyday language rather than probability expressions. They also pointed out that counterfactual explanations and SHAP-based representations can offer meaningful insights when presented as key contributing factors with their relative influence, though both can introduce additional system complexity and computation cost, particularly in dynamic or resource-sensitive environments.

There was a clear and notable convergence between the interview findings and the survey results, despite the difference in participant roles and levels of expertise. While the interview questions were primarily oriented towards technical aspects of explainability, including the challenges involved in developing explanatory mechanisms, approaches for adapting explanations for different types of users, the trustworthiness of explanation techniques, and the scalability of a multi-stage explanatory pipeline, the perspectives expressed by experts aligned closely with the expectations and preferences reported by potential end users in the survey.

Both data sources highlighted clarity and brevity as essential for successful explanation. The survey reflected a clear preference for explanations that begin with a concise summary, while interview participants emphasised that lengthy or overly detailed explanations can impede understanding and increase cognitive effort.

The integration of textual explanations with visual elements emerged as another shared preference. The survey indicated favourability toward explanations supported by visual cues, such as highlighting or simple graphical markers, while interview participants noted that visual support helps structure information and enhance its accessibility for both expert and non-expert users.

Furthermore, both the survey and interview findings underscored the importance of responsiveness and computational efficiency. Slow system responses or high processing demands were viewed as potential barriers to usability, and interview participants cautioned that some explanation techniques with high computational cost may be unsuitable for real-time interaction.

Finally, both sources affirmed that transparency and traceability are central to building trust. The survey revealed strong value placed on being able to understand the reason behind classification outcomes, while interview participants emphasised the need for explanations to reflect the genuine reasoning of the model rather than presenting superficially convincing justifications.

## VI. Discussion

### A. Survey

The survey depicts a user base that is digitally literate and largely positive about an explainable sentiment-analysis tool for healthcare, but with clear expectations around clarity, trustworthiness, and speed. Read through the Technology Acceptance Model (TAM), adoption hinges on two levers: perceived usefulness (does it save time, surface what matters, and fit existing review-reading workflows?) and perceived ease of use (are explanations concise and accessible. If explanation delivery is slow or cognitively heavy, both usefulness and the intention to use decline, exactly as TAM predicts [5].

Findings on explainability and trust suggest that transparency is not optional: respondents want to understand why classifications were made and showed higher trust when explanations are available. This accords with work on trust calibration in automation, where appropriate transparency helps users avoid both over reliance and under reliance, and with user-centred views of explanation that emphasise communicative clarity over technical exhaustiveness.

Preferred formats point to a layered, text-first design with optional visuals: brief plain-language rationales for quick scanning, drill-down to salient phrases or aspects, and (where helpful) compact graphics. This aligns with the XAI principles articulated by the National Institute of Standards and Technology (NIST), which state that explanations should be meaningful to the intended user and faithful to the model's underlying reasoning, rather than generic or uniformly applied across users [17]. This underscores the value of an explanation structure that begins with a concise core rationale while allowing additional detail when required.

Thematic responses on adoption enablers and barriers reinforce these implications. Enablers include usability and simplicity (low-friction UI; minimal learning effort), accuracy and reliability (clear evidence for outputs; no misleading simplifications), and speed/efficiency (near-instant responses that genuinely reduce reading time). Barriers include complex or overly long explanations, opacity (not knowing how outputs were derived or being unable to trace to the source text), data credibility (fake/marketing reviews), and privacy/security concerns.

### B. Interview

The interviews converge on three practical priorities for explainable AI in healthcare reviews: clarity and audience-fit, faithfulness of the explanation to the underlying model, and performance/operational viability. Viewed through the Technology Acceptance Model (TAM), these priorities directly map onto perceived ease of use (clear, succinct, jargon-light explanations that match the user's expertise) and perceived usefulness (explanations that genuinely aid judgement and fit existing workflows without introducing delays) [5].

A key insight from the interviews is that layered explanations are not only beneficial for usability, but also essential for maintaining fidelity to the underlying model. Experts emphasised that starting with a concise, plain-language rationale is appropriate, provided that users are able to access the underlying reasoning when needed. The concern was not simplification itself, but oversimplification that risks misrepresenting model behaviour. Accordingly, the layered structure is viewed as a mechanism to balance accessibility and faithfulness: the initial explanation communicates the core decision logic, while optional deeper layers preserve transparency and support verification. This reflects contemporary XAI perspectives that explanations should be tailored to user context while remaining grounded in the actual decision-making process [17].

Regarding trust, the interviews highlight the importance of transparency. Participants emphasised the need for faithful explanations rather than superficially convincing explanations, and advocated the integration of complementary methods such as counterfactual "what-if" scenarios to enhance actionability, local feature attributions (e.g., SHAP) to provide principled importance scores, and simple local surrogate models to deliver quick, human-readable sketches. Such evidence indicates that carefully designed transparency mechanisms can foster appropriate reliance on automation, avoiding both over trust and under trust [6].

At the method level, the interviews endorse a toolkit rather than a single canon: Counterfactuals are valued for being intuitive and actionable ("what would need to change to obtain a different outcome?") [18]. SHAP is valued for its additive, consistency-based formulation that unifies several attribution methods and offers a strong theoretical footing for local importance [19]. Simple local surrogates can provide fast, human-readable logic if used with care to preserve local fidelity (the interviews' fidelity concerns reflect this caution). These preferences are broadly consistent with recent XAI literature.

The interviews also surface implementation constraints that matter for real-world adoption. Post-hoc explainers (e.g., SHAP/LIME) and multi-layer pipelines may introduce computational overhead, latency, and potential instability at scale, especially with large-scale datasets. Several responses therefore point towards performance-aware design patterns: pre-computation and caching where feasible; prudent approximation (that is, using simplified yet reliable estimates to reduce computational cost without substantially compromising accuracy); staged or interactive delivery that yields incremental insight quickly (reducing perceived wait time); and human-in-the-loop review for cases that require human oversight. These patterns preserve users' sense of utility and flow, again reinforcing TAM's contention that usefulness and ease jointly drive acceptance [5]. These participant concerns directly inform the system's design, highlighting the need for lightweight, efficient, and incrementally delivered explanations.

## VII. Future Work

These combined survey and interview findings provide a clear direction for the next steps in designing and evaluating the system.

A layered, audience-sensitive explanation framework will be developed to account for the differing needs of user groups within the prototype. The system will initially present a concise, plain-language explanation summarising the model's decision. Users can then optionally expand the explanation to view the specific segments or aspects that influenced the classification. Visual elements, such as colour gradients, simple icons, or contribution indicators, are included as supportive aids to enhance comprehension for users with varying levels of familiarity. Where deeper examination is required, more detailed representations such as ranked feature importance can be accessed. To ensure smooth responsiveness and avoid user fatigue, explanation components will be delivered incrementally, with the core textual explanation displayed first and additional layers retrieved only when requested. This structure supports both novice and experienced users by enabling low-effort reading when desired while also providing interpretable and verifiable depth when needed. It reduces cognitive load and strengthens users' ability to rely on the system by aligning the explanation with the level of detail required, without compromising accuracy or transparency.

The next stage of the research focuses on implementing this layered explanatory approach and evaluating it within a functional prototype. Initial A/B testing will compare (a) a brief text-only rationale and (b) a text rationale with highlighted spans, with optional visual indicators available only when requested. The evaluation will assess the effect of the explanations on user understanding, perceived trust, cognitive effort and willingness to rely on the system in real healthcare review contexts.

## VIII. Limitations

This study deliberately focused on a digitally literate, highly educated participant group, providing valuable early insights into adoption and trust patterns among likely early adopters of XAI systems. While this focus strengthens the internal validity of our findings, it also limits the generalisability of the results to patients with lower levels of digital literacy or limited familiarity with AI-based systems. Future work will therefore intentionally recruit participants from more diverse backgrounds, including individuals with less experience using digital platforms. This recruitment will broaden digital literacy levels and ensure diversity across age groups and everyday healthcare engagement contexts, allowing explanation preferences to be evaluated among users with different technological familiarity, cognitive needs, and patterns of interaction with healthcare services. Similarly, the expert interview sample was intentionally small (N = 5) to enable in-depth exploration of technical and methodological issues; future studies may build on these insights with larger and more representative expert panels.

.

## IX. Recommendations

The findings from both the survey and interviews converge on a set of actionable recommendations to guide the design and deployment of explainable AI systems for healthcare reviews. **First,** explanations should be clear, contextually appropriate, and tailored to audience expertise, ensuring that technical users receive sufficient detail while non-technical stakeholders benefit from simplified forms. **Second,** strong fidelity safeguards are essential to guarantee that explanations reflect the model's actual reasoning rather than superficial plausibility. This includes the use of principled attribution methods and evidence highlighting whenever computational budgets permit. **Third,** explanations must be delivered with performance-aware mechanisms, such as pre-computation, caching, or progressive rendering, in order to minimise latency and preserve user engagement. **Fourth,** a mixed-methods explanation toolbox should be adopted, combining counterfactual scenarios for actionability, attribution-based methods for principled importance scores, and faithful surrogate models for fast, human-readable sketches. **Fifth,** provenance should be built in from the design stage to ensure transparent explanations and system trust. Privacy safeguards should also be included to protect user data and meet ethical and legal standards.

## X. Conclusion

Explainable AI systems are becoming an increasingly essential component in healthcare. However, current approaches face constraints related to clarity, fidelity, and performance. Our study shows that there is strong potential for adoption when a careful balance is achieved across these dimensions. One major challenge is that no single explanation method is sufficient to meet the needs of all users. This motivated us to combine two studies, a survey and interviews, involving participants with different levels of expertise. The survey results indicate a clear demand for explanations that are both understandable and faithful, delivered within acceptable performance limits. The interviews further confirmed the importance of providing layered, audience-tailored explanations, together with fidelity safeguards and performance-aware design patterns. In this context, our future work will focus on translating these findings into practice through the development of a prototype system that integrates multiple explanation methods with performance-aware delivery and built-in privacy safeguards. Such a system will enable the evaluation and validation of explainable AI applications in healthcare before their wider deployment in real-world decision-making environments.